\documentclass[prd,a4paper,preprint,preprintnumbers,nofootinbib,superscriptaddress]{revtex4-2}
\usepackage[a4paper, hdivide={1.91cm,,1.165cm}, vdivide={1.83cm,,3.0cm}]{geometry}

\usepackage[hyperfootnotes=false]{hyperref}
\hypersetup{linkcolor=red,
citecolor=green,
filecolor=cyan,
urlcolor=magenta}

\usepackage{amsfonts}

\usepackage{amsmath}
\usepackage{amssymb}

\usepackage{bbm}

\usepackage{bbold}

\usepackage{adjustbox}
\usepackage{booktabs}

\usepackage{colortbl}

\usepackage{cancel}

\usepackage{color}
\usepackage{xcolor}

\usepackage{graphicx}
\usepackage{xspace}

\usepackage{units}

\usepackage{slashed}

\usepackage{tensor}

\usepackage{gensymb}

\usepackage{cleveref}

\usepackage{tabularx}
\usepackage{multirow}

\usepackage{setspace}
\usepackage{orcidlink}

\usepackage{caption}
\usepackage{subcaption}

\usepackage{enumerate}
\usepackage{enumitem}

\usepackage[version=3]{mhchem}

\usepackage{lipsum}

\newcommand{{\ia} }{{\i}}
\newcommand{{\Ia} }{{\.I}}

\newcommand{\ra}{\rightarrow}
\newcommand{\la}{\leftarrow}

\def\s2tw{{\rm sin ^2 \theta_{W}}}

\def\beq{\begin{equation}}
\def\eeq{\end{equation}}
\def\bea{\begin{eqnarray}}
\def\eea{\end{eqnarray}}

\def\vel{\left|}
\def\ver{\right|}
\def\nnb{\nonumber}
\def\lla{\left<}
\def\rra{\right>}
\def\nnb{\nonumber}
\def\la{\langle}
\def\ra{\rangle}

\def\qq{\la \bar q q \ra}

\def\es{ &=& }

\def\ek{&-& }
\def\cp{&\times&}

\def\qq{\langle \bar q q \rangle}

\def\bea{\begin{eqnarray}}
\def\eea{\end{eqnarray}}
\def\beeq{\begin{eqnarray}}
\def\eeeq{\end{eqnarray}}

\def\vel{\left|}
\def\ver{\right|}
\def\nnb{\nonumber}

\def\lla{\left<}
\def\rra{\right>}

\def\nnb{\nonumber}
\def\la{\langle}
\def\ra{\rangle}
\def\ba{\begin{array}}
\def\ea{\end{array}}

\def\xis0{{\Xi^{*0}}}

\def\qq{\la \bar q q \ra}

\def\g5{\gamma_5}

\def\es{\!\!\! &=& \!\!\!}

\def\ek{&-& \!\!\!}

\def\cp{&\times& \!\!\!}

\allowdisplaybreaks[1]

\begin{document}


\title{Analysis of the strong decays of SU(3) partners of the $\Omega(2012)$ baryon}

\author{T.~M.~Aliev\,\orcidlink{0000-0001-8400-7370}}
\email{taliev@metu.edu.tr}
\affiliation{Department of Physics, Middle East Technical University, Ankara, 06800, Turkey}

\author{S.~Bilmis\,\orcidlink{0000-0002-0830-8873}}
\email{sbilmis@metu.edu.tr}
\affiliation{Department of Physics, Middle East Technical University, Ankara, 06800, Turkey}
\affiliation{TUBITAK ULAKBIM, Ankara, 06510, Turkey}

\author{M.~Savci\,\orcidlink{0000-0002-6221-4595}}
\email{savci@metu.edu.tr}
\affiliation{Department of Physics, Middle East Technical University, Ankara, 06800, Turkey}

\date{\today}

\begin{abstract}
  We estimate the coupling constants and decay widths of the $SU(3)$ partners of the $\Omega(2012)$ hyperon, as discovered by the BELLE Collaboration, using the light cone sum rules method. Our study includes a comparison of the obtained results for relevant decay widths with those derived within the framework of the flavor $SU(3)$ analysis. We observe a good agreement between the predictions of both approaches. The results we obtain for the branching ratio can provide helpful insights for determining the nature of the $SU(3)$ partners of the $\Omega(2012)$ baryon.
\end{abstract}

\maketitle


\newpage



\section{Introduction\label{intro}}
In 2018, the BELLE Collaboration made an exciting announcement regarding the discovery of the  $\Omega(2012)$ hyperon. This discovery was based on the  $\Omega^{*-} \to \Xi^0 K^-$ and $\Omega^{*-} \to \Xi^- K_s^0$ decay channels, with a measured mass of $m = 2012.4 \pm 0.7 \,{\rm(stat)} \pm 0.6 \,{\rm(sys)}\,\rm{MeV}$, and decay width of $\Gamma_{tot} = 6.4_{-2.0}^{+2.5} \, {\rm(stat)} \pm 1.6 \,{\rm(sys)}\,\rm{MeV}$ \cite{Belle:2018mqs}. However, knowing only the mass of the state is not sufficient enough to determine the quantum numbers of a state. For instance, within the QCD sum rule method, the mass of the  $\Omega(2012)$ baryon is estimated, assuming it to be either 1P or 2S excitation state \cite{Aliev:2016jnp}. Both assumptions yield the same mass value, although the estimated residues differ. Thus, additional physical quantities, such as the decay width, are necessary to identify the quantum numbers of newly discovered particles.

In a previous study \cite{Aliev:2018yjo}, the $\Omega(2012) \to \Xi^0 K^-$ transition was investigated, and its corresponding decay width was estimated by considering two possible
scenarios for $\Omega(2012)$: either a $1P$ or $2S$ state. A comparison of the total decay widths obtained in this work led to the conclusion that the  $\Omega(2012)$ is itself a $J^P = {3\over2}^-$ state. Moreover, predictions from various theoretical models also converge on the likely  quantum numbers $J^P = \frac{3}{2}^-$ for the observed state \cite{Faustov:2015eba, CLQCD:2015bgi, An:2014lga,An:2013zoa,Engel:2013ig,
  Pervin:2007wa,Liu:2007yi,Oh:2007cr,Loring:2001ky,Capstick:1986ter,Chao:1980em,Kalman:1982ut}.

In this study, considering $\Omega(2012)$ as $J^P = \frac{3}{2}^-$ state, strong couplings of SU(3) partners of this state are investigated within the framework of light cone sum rules (LCSR). It should be noted that this problem was also studied in \cite{Polyakov:2018mow} using the flavor SU(3) symmetry approach.

The structure of this paper is as follows: Section~\ref{sec:2} introduces the LCSR for the strong couplings of the transitions  ${3\over2}^- \to  {1\over2}^+ +  \text{pseudoscalar mesons}$. Section~\ref{sec:3} provides a numerical analysis of the LCSR, focusing on the relevant strong couplings. Within this section, we also present the computed values of the decay widths based on the obtained coupling constants. Additionally, we compare our results with those obtained from the flavor SU(3) symmetry method. Finally, our conclusions are summarized in the last section.
\section{LCSR for the strong couplings of $SU(3)$ partners of $\Omega(2012)$}
\label{sec:2}
To calculate the strong couplings of $SU(3)$ partners, denoted as ${3 \over 2}^-$ states in the following discussions, we introduce
the vacuum-to-octet baryon correlation function:
\bea
\label{eq1}
\Pi_{\mu\nu}(p,q) = i \int d^4x e^{iqx} \lla 0 \vel T\Big\{ \eta_\mu(0)
J_\nu(x) \Big\} \ver {\cal O}(p) \rra~,
\eea
where $\eta_\mu$ represents the interpolating current of the decuplet baryons,
$J_\nu = \bar{q}_1 \gamma_\nu \gamma_5 q_2$ is the interpolating current of
the pseudoscalar mesons, and $\vel  {\cal O}(p) \rra$ represents the octet baryon
state.

The interpolating current of the decuplet baryons can be written as:
\bea
\label{eq2}
\eta_\mu \es \varepsilon^{abc} A \Big\{ \left(q_1^{aT} C \gamma_\mu q_2^b\right)
q_3^c + \left(q_2^{aT} C \gamma_\mu q_3^b \right) q_1^c +
\left(q_3^{aT} C \gamma_\mu q_1^b \right) q_2^c \Big\}~,
\eea
where $a,b,c$ are the color indices, $C$ is the charge conjugation operator,
and $A$ is the normalization factor. The quark content of the decuplet baryons and the normalization
factor $A$ are presented in Table \ref{tab1}. 

%
\begin{table}[hbt]
\begin{adjustbox}{center}
\renewcommand{\arraystretch}{1.2}
\setlength{\tabcolsep}{6pt}
  \begin{tabular}{lcccc}
    \toprule
                   &         $A$          & $q_1$ & $q_2$ & $q_3$  \\
\midrule
$\Delta^+$         &   $\sqrt{1\over 3}$ &  $u$  &  $u$  & $d$    \\
$\Sigma^{\ast +}$  &   $\sqrt{1\over 3}$  &  $u$  &  $u$  & $s$    \\
$\Sigma^{\ast 0}$  &   $\sqrt{2\over 3}$  &  $u$  &  $d$  & $s$    \\
$\Sigma^{\ast -}$  &   $\sqrt{1\over 3}$  &  $d$  &  $d$  & $s$    \\
$\Xi^{\ast 0}$     &   $\sqrt{1\over 3}$  &  $s$  &  $s$  & $u$    \\
$\Xi^{\ast -}$     &   $\sqrt{1\over 3}$  &  $s$  &  $s$  & $d$    \\
    \bottomrule
  \end{tabular}
\end{adjustbox}
\caption{The quark content of the decuplet baryons and the normalization
factor $A$.}
\label{tab1}
\end{table}
%

To derive the Light Cone Sum Rules (LCSR) for the strong coupling constants, the approach involves computing the correlation function in two ways: in terms of hadrons and in terms of quark-gluon fields within the deep Euclidean domain. By applying the quark-hadron duality ansatz, the relevant sum rules can be derived.

The calculation of the strong coupling constants in the framework of the LCSR is
based on the fact that they appear in double dispersion relation for the
same correlation function given in Eq. (\ref{eq1}). In other words, calculating the
strong coupling constant requires the use of double dispersion relation for
the correlation function by making use of the axial vector current.

Before delving into the details of the calculations, it is important to highlight the following aspect: the interpolating current for the decuplet baryons
interacts not only with the ground positive parity states $J^P = {3\over2}^+$
 but also with the negative parity states $J^P = {3\over2}^-$  and even with states of
$J^P = {1\over2}^-$. 

To eliminate the contributions from unwanted states $J^P = {3\over2}^+$
and $J^P = {1\over2}^-$, a technique involving linear contributions of  different
Lorentz structures is employed (for more details about this approach refer to \cite{Khodjamirian:2011jp}).

Following the standard procedure, we insert the total set of 
set of baryons with $J^P = {3\over2}^+$ into the correlation function as well
as the corresponding pseudoscalar mesons. Then, we get,
\bea
\label{eq3}
\Pi_{\mu\nu} (p,q) \es \sum_{i=\pm}
{ \lla 0 \vel \eta_\mu \ver {3\over 2}^i (p^\prime) \rra \over m_i^2 - p^{\prime 2} }
\,{ \lla {3\over 2}^i  (p^\prime)  {\cal P}(q) | {\cal O}(p) \rra \over
m_{\cal P}^2 - q^2 }
\,\lla 0 \vel J_\nu(x) \ver {\cal P}(q) \rra~,
\eea
where summation is over positive and negative states, and $m_{\cal P}$ is the mass of
the corresponding pseudoscalar meson. The matrix elements in the above equation 
are defined as,

\bea
\label{eq4}
\lla 0 \Big\vert \eta_\mu \Big\vert {3\over 2}^+ (p^\prime) \rra \es \lambda_+ 
u_\mu (p^\prime)~,\nnb \\
\lla 0 \Big\vert \eta_\mu \Big\vert {3\over 2}^- (p^\prime) \rra \es \lambda_-
\gamma_5 u_\mu (p^\prime)~,\nnb \\
\lla {3\over 2}^+ (p^\prime) {\cal P}(q)   \Big\vert  {\cal O} (p) \rra \es g_+
\bar{u}_\alpha (p^\prime) \gamma_5 u(p) q^\alpha~,\nnb \\
\lla {3\over 2}^- (p^\prime) {\cal P}(q)   \Big\vert  {\cal O} (p) \rra \es g_-
\bar{u}_\alpha (p^\prime) u(p) q^\alpha~,\nnb \\
\lla 0 \vel J_\nu \ver {\cal P}(q) \rra \es i f_{\cal P} q_\nu~,
\eea
where $\lambda_\pm$ are the residues of the related ${3\over 2}^\pm$
baryons, $g_\pm$ stands for the coupling constants of the 
$J^P = {3\over2}^\pm$ baryons with the octet baryons and the pseudoscalar mesons, 
$f_{\cal P}$ is the decay constant of the pseudoscalar meson
and $q$ denotes its 4-momentum, and $u_{\mu}(p^\prime)$ and $u(p)$ are the Rarita-Schwinger and Dirac spinors respectively. Performing summation over the spins of the 
Rarita-Schwinger spinors with the help of the following formula,
\bea
\label{eq5}
\sum_{s^\prime} u_\mu(p^\prime,s^\prime) \bar{u}_\alpha(p^\prime,s^\prime)
= - (\slashed{p}^\prime + m) \Bigg[ g_{\mu\alpha} - {1\over 3} \gamma_\mu
\gamma_\nu - {2 p_\mu^\prime p_\alpha^\prime \over 
 3 m^2} +{p_\mu^\prime \gamma_\alpha - p_\alpha^\prime \gamma_\mu \over 3 m}
\Bigg]~,
\eea
and using Eqs. (\ref{eq3}) and (\ref{eq4}) one can obtain the expression of
the correlation function from the hadronic part.
It should be reminded here that the interpolating current interacts not only with spin ${3\over 2}$ states,
but also with spin ${1\over 2}$ states.

Using the condition $\gamma^\mu \eta_\mu=0$, it can easily be shown that
\bea
\label{eq6}
\lla 0 \Big\vert \eta_\mu \Big\vert {1\over 2} (p^\prime) \rra \sim
\left[ \alpha \gamma_\mu - \beta p_\mu^\prime \right] u(p^\prime)~.
\eea

It follows from this equation that any structure containing $\gamma_\mu$
or $p_\mu^\prime$ is ``contaminated" by the contributions of spin ${1\over 2}$-states. Hence, to remove the contributions of spin ${1\over2}$-states, these structures are all discarded. Another problem is all Dirac structures not being independent of each other.
To overcome this issue, Dirac structures need to be arranged in a specific order. In the present work we choose the ordering $\gamma_\mu \slashed{p}^\prime
\slashed{q} \gamma_\nu \gamma_5$.

Keeping these notes in mind, and using Eqs. (\ref{eq3}),
(\ref{eq4}) and (\ref{eq5}), we obtain the correlation function from the phenomenological part as follows:
\bea
\label{eq7}
\Pi_{\mu\nu} \es {\lambda_+ g_+ (- \slashed{q} + m_+ - m_{\cal O}) \gamma_5 q_\mu
q_\nu f_{\cal P} \over (m_+^2 - p^{\prime 2}) (m_{\cal P}^2 - q^2 )} u(p) +
{\lambda_- g_- ( \slashed{q}+ m_- + m_{\cal O}) \gamma_5 q_\mu
q_\nu f_{\cal P} \over (m_-^2 - p^{\prime 2}) (m_{\cal P}^2 - q^2 )} u(p)~,
\eea
where $m_{\cal O}$ is the mass of the relevant octet baryon, $m_+(m_-)$ is
the mass of the spin-${3\over 2}$  positive (negative)
parity baryon, respectively.

As a last step, we need to eliminate the contributions of $J^P = \frac{3}{2}^+$ states. For this purpose, we use the linear combinations of the invariant functions corresponding to different Lorentz structures.

We now turn our attention to the calculation of the correlation function by
using the operator product expansion (OPE) in the deep Euclidean region for
the variables $p^{\prime 2} = (p-q)^2$, and $q^2 \ll 0$. The new element of
the calculation is the appearance of the double spectral density of the invariant functions.
For the calculation of the double spectral densities it is enough to find
the double spectral representations of the master integrals of the form,
\bea
\label{nolabel}
I_{n,k} = \int du {u^k\over \left[m^2 - (p u - q)^2\right]^n}~;
~~n=1,2,3. \nnb
\eea

We now present the details of the calculations for the spectral density for
$n=1$ case. The cases $n=2$ and $n=3$ are calculated in the similar manner.
First of all we will show how the doubly spectral density can be obtained
from the invariant amplitudes.
The invariant amplitudes can be written in terms of the double spectral
representation as follows
\bea
\label{eq8}
\Pi[(p-q)^2,q^2] = \int ds_1\int ds_2 {\rho(s_1,s_2) \over [s_1-(p-q)^2]
(s_2-q^2)} + \cdots
\eea
The spectral density can be obtained from $\Pi[(p-q)^2,q^2]$ by applying two
subsequent double Borel transformations. After first double Borel
transformation over the
variables $-(p-q)^2$ and $-q^2$ we get,
\bea
\label{nolabel}
\Pi^{{\cal B}_1} (M_1^2,M_2^2) = \int ds_1 \int ds_2 e^{-s_1/M_1^2
-s_2/M_2^2} \rho(s_1,s_2)~.
\eea
Before implementing second double Borel transformation, we introduce new
variables $\sigma_1 = {1\over M_i^2}$. The second double Borel
transformation can be  performed over the new Borel parameter $\tau_i$ by using the
relation,
\bea
\label{nolabel}
{\cal B}_\tau e^{-s\sigma} = \delta\left({1\over\tau}-s\right)~. 
\eea
As a result we get
\bea
\label{nolabel}
{\cal B}_{\tau_1} {\cal B}_{\tau_2} \Pi^{{\cal B}_1}(M_1^2,M_2^2) =
\rho\left({1\over\tau_1},{1\over \tau_2}\right)~.
\eea
Hence, double spectral density can be obtained as follows,
\bea
\label{nolabel}
\rho(s_1,s_2) = {\cal B}_{1\over s_1}(\sigma_1) {\cal B}_{1\over
s_2}(\sigma_2) \Pi^{\cal B}
\left({1\over \sigma_1},{1\over \sigma_2}\right)~. \nnb
\eea 

Let us now pay our attention to the double spectral density for the $n=1$ 
case.  Using
\bea
\label{nolabel}
 - (pu - q)^2 = -u (p-q)^2 - \bar{u} q^2 + u \bar{u} m_{\cal O}^2~,\nnb
\eea
where $\bar{u}=1-u$. $I_{1,k}$ can be written as,
\bea
\label{nolabel}
I_{1,k} \es \int du {u^k \over \left[m^2 - u (p-q)^2 - \bar{u} q^2 +
\bar{u}um_{\cal O}^2 \right]} \nnb \\
\es \int du {u^k\over {\cal D}}~,\nnb
\eea
where $m$ is the corresponding quark mass.
Using the Schwinger representation for the denominator and carrying out the
first double Borel transformation over the variables $-(p-q)^2$ and $-q^2$,
we get
\bea
\label{nolabel}
I_{1,k} \es {\sigma_2^k \over (\sigma_1 + \sigma_2)^{k+1}} \, exp \left[-m_{\cal O}^2
{\sigma_1 \sigma_2 \over \sigma_1 + \sigma_2} - m^2 (\sigma_1 +
\sigma_2)\right]~,\nnb \\
\es {\sigma_2^k \over (\sigma_1 + \sigma_2)^{k+1}} \, exp \left[m_{\cal O}^2 
{\sigma_1^2 +  \sigma_2^2 \over 2(\sigma_1 + \sigma_2)} -
\left(m^2 + {m_{\cal O}^2 \over 2}\right) (\sigma_1 + \sigma_2) \right]~,
\nnb
\eea
where $\sigma_i = \frac{1}{M_i^2}$.
In order to perform the second double Borel transformation we use the
relation,
\bea
\label{nolabel}
\sqrt{\sigma_1 + \sigma_2 \over 2 \pi} \int_{-\infty}^{+\infty} dx_i 
\, exp \left[{-{\sigma_1 + \sigma_2 \over 2} x_i^2 - \sigma_i m_{\cal O}
x_i}\right] = 
\, exp \left[{ m_{\cal O}^2 \sigma_i^2 \over 2(\sigma_1 + \sigma_2)}\right]~.
\nnb
\eea
Then we get,
\bea
\label{nolabel}
I_{1,k}^{\cal B} \es {1\over 2 \pi} \int_{-\infty}^{+\infty} dx_1
\int_{-\infty}^{+\infty} dx_2 \, 
{\sigma_2^k \over (\sigma_1 + \sigma_2)^k} 
\, exp \Bigg[- \sigma_1 \Bigg(m^2 + {(m_{\cal O} + x_1)^2 + x_2^2 \over 2}\Bigg) \nnb \\
\ek \sigma_2 \Bigg(m^2 + {(m_{\cal O} + x_2)^2 + x_1^2 \over
2}\Bigg)\Bigg]~,\nnb \\
\es {1\over 2 \pi} {1\over \Gamma(k)} \int_{-\infty}^{+\infty} dx_1
\int_{-\infty}^{+\infty} dx_2\int_{0}^{\infty} dt\, t^{k-1} \sigma_2^k
\, exp\Bigg[- \sigma_1 \Bigg(m^2 + {(m_{\cal O} + x_1)^2 + x_2^2 \over 2} + t \Bigg) \nnb \\
\ek \sigma_2 \Bigg(m^2 + {(m_{\cal O} + x_2)^2 + x_1^2 \over
2} + t \Bigg)\Bigg]~,\nnb \\
\es {1\over 2 \pi} {1\over \Gamma(k)} \int_{-\infty}^{+\infty} dx_1
\int_{-\infty}^{+\infty} dx_2\int_0^{\infty} dt\, t^{k-1}
\, exp \Bigg[- \sigma_1 \Bigg(m^2 + {(m_{\cal O} + x_1)^2 + x_2^2 \over 2} + t \Bigg) \Bigg] \nnb \\
\cp \Bigg( - {\partial \over \partial t}\Bigg)^k  \, exp \Bigg[ - \sigma_2 \Bigg(m^2 + {(m_{\cal O} + x_2)^2 + x_1^2 \over
2} + t \Bigg)\Bigg]~.\nnb
\eea
After performing the second Borel transformation, we obtain the the spectral
density corresponding to $I_{1,k}$ as is given below,
\bea
\label{nolabel}
\rho_{1,k}(s_1,s_2) \es {1\over 2 \pi} {1\over \Gamma(k)} \Bigg(-\frac{\partial}{\partial s_2} \Bigg)^k \int_{-\infty}^{+\infty} dx_1
\int_{-\infty}^{+\infty} dx_2\int_0^{\infty} dt\, t^{k-1}
\delta\Bigg[ s_1 - \Bigg(m^2 + {(m_{\cal O} + x_1)^2 + x_2^2 \over 2} + t \Bigg)\Bigg]\nnb \\
\cp \delta\Bigg[ s_2 - \Bigg(m^2 + {(m_{\cal O} + x_2)^2 + x_1^2 \over 2} + t \Bigg)\Bigg]\nnb \\
\es {1\over 2 \pi} {1\over \Gamma(k)} \Bigg( - {\partial \over \partial s_2}\Bigg)^k 
\int_{-\infty}^{+\infty} dx_1 \int_{-\infty}^{+\infty} dx_2\int_0^{\infty} dt\, t^{k-1}
\delta\Bigg[ s_1 - \Bigg(m^2 + {(m_{\cal O} + x_1)^2 + x_2^2 \over 2} + t \Bigg)\Bigg]\nnb \\
\cp \delta\Bigg[ s_2 - \Bigg(m^2 + {(m_{\cal O} + x_2)^2 + x_1^2 \over 2} + t
\Bigg)\Bigg]\nnb~.
\eea
Using two Dirac delta functions, one can easily perform integrals over
$t$ and $x_2$ whose result is given below,
\bea
\label{nolabel}
\rho_{1,k}(s_1,s_2) \es {1\over 2 \pi  \Gamma(k) m_{\cal O}} \Bigg( - {\partial \over \partial s_2}\Bigg)^k \int_{-\infty}^{+\infty} dx_1
  \Bigg[s_1 - \Bigg(m^2 + {(m_{\cal O} + x_1)^2 + x_2^2 \over 2} \Bigg)\Bigg]^{k-1} \nnb \\
\cp \Theta\Bigg[s_1 - \Bigg(m^2 + {(m_{\cal O} + x_1)^2 + x_2^2 \over 2} \Bigg)\Bigg]~, \nnb
\eea
where
\bea
\label{nolabel}
x_2 = {s_2 - s_1 \over m_{\cal O}} + x_1~,\nnb
\eea
and $\Theta(x)$ is the Heaviside step function which restricts the integral
over $x_1$ between the limits $y_\pm(s_1,s_2)$ where
\bea
\label{nolabel}
y_\pm(s_1,s_2) = {- m_{\cal O}^2 + s_1 - s_2 \pm \sqrt{\Delta} \over 2 m_{\cal O}}~,\nnb
\eea
and
\bea
\label{nolabel}
\Delta =  - m_{\cal O}^4 - (s_1-s_2)^2 + 2 m_{\cal O}^2 (-2 m^2 +s_1+s_2)~.\nnb
\eea
Thus as a result of above summarized calculations, the spectral density can
takes the following form,
\bea
\label{nolabel}
\rho_{1,k}(s_1,s_2) \es {1\over 2 \pi} {1\over \Gamma(k)} {1\over m_{\cal O}}
\Bigg( - {\partial \over \partial s_2}\Bigg)^k
\int_{y_-}^{y_+}dx \Big[ (y_+-x) (x-y_-)\Big]^k \Theta(\Delta)~. \nnb
\eea
In order to evaluate the $x$ integral, we introduce a new variable through
the relation,
\bea
\label{nolabel}
x = (y_+-y_-) y +y_-~,\nnb
\eea
so that the  spectral density can be written as,
\bea
\label{eq9}
\rho_{1,k}(s_1,s_2) \es {1\over 2 \pi} {\Gamma(k)\over \Gamma(2k)} {1\over m_{\cal
O}^{2 k}} \Bigg( - {\partial \over \partial s_2}\Bigg)^k
\Big[\Delta^{k-{1\over 2}}\Theta(\Delta) \Big]~.
\eea

Double spectral densities for $I_{2,k}$ and $I_{3,k}$ can be calculated with
the help of the following relations,
\bea
\label{nolabel}
I_{2,k} \es \Bigg(-{\partial\over \partial m^2} \Bigg)I_{1,k}~,~\mbox{and,}\nnb \\
I_{3,k} \es {1\over 2} \Bigg(-{\partial\over \partial m^2}\Bigg)^2 I_{1,k}~,\nnb
\eea
(see also \cite{Khodjamirian:2011jp} for the calculation of the spectral densities $I_{2,k}$ and $I_{3,k}$).

Matching the OPE results with the double dispersion relations for the
relevant Lorentz structures for the hadrons, applying the quark-hadron
duality ansatz, and performing double Borel transformation with respect to
the variables $-(p-q)^2$ and $-q^2$, we obtain the LCSR for the relevant
coupling constants whose explicit form can be written as,
\bea
\label{eq10}
g_- \es { e^{m_-^2/M_1^2} \,e^{m_{\cal P}^2/M_2^2}\over f_{\cal P} \lambda_-(m_+
+m_-) } {1\over \pi^2} \int_0^{s_0} ds_1 \nnb \\
\cp \int_{t_1(s_1)}^{t_2(s_1)}
ds_2 \, e^{-s_1/M_1^2} e^{-s_2/M_2^2} \, {\mbox Im}_{s_1} {\mbox Im}_{s_2}
\Big\{\Pi_1 (m_+ - m_{\cal O}) + \Pi_2 \Big\}~, 
\eea
where $\Pi_1$ and $\Pi_2$ are the invariant functions of the Lorentz
structures $\slashed{q} \gamma_5 q_\mu q_\nu$ and $\gamma_5 q_\mu q_\nu$,
respectively, and
\bea
\label{nolabel}
t_{1,2} = s_1 + m_{\cal O}^2 \mp 2 m_{\cal O} \sqrt{s_1-m^2}~.\nnb
\eea
\section{Numerical Analysis}
\label{sec:3}
The present section is devoted to the numerical analysis of the coupling constants derived in the previous section within LCSR.
The main nonperturbative input of the considered LCSR is the distribution amplitudes (DAs) of the octet baryons, namely $N$, $\Sigma$ and $\Xi$. The explicit expressions of the relevant DAs as well as the values of the parameters ($f$, $\lambda_1$, and $\lambda_2$) determined from the analysis of mass sum rules~\cite{Braun:2000kw,Liu:2008yg,Liu:2009uc,Liu:2010bh} are presented in Table~\ref{tab:constants} for completeness.
\begin{table}[hbt]
\renewcommand{\arraystretch}{1.2}
\setlength{\tabcolsep}{6pt}
  \begin{tabular}{lccc}
    \toprule
                   &         $f~(\rm{GeV^2}) $          & $\lambda_1~(\rm{GeV^2})$ & $\lambda_2~(\rm{GeV^2})$   \\
\midrule
N  & $(5.3 \pm 0.5) \times 10^{-3} $ & $- (2.7 \pm 0.9) \times 10^{-2}  $ & $(5.1 \pm 1.9) \times 10^{-2}  $ \\
$\Sigma$  & $(9.4 \pm 0.4) \times 10^{-3}  $  & $- (2.5 \pm 0.1) \times 10^{-2}  $ & $(4.4 \pm 0.1) \times 10^{-2}  $ \\
$\Xi$  & $(9.9 \pm 0.4) \times 10^{-3}  $ & $- (2.8 \pm 0.1) \times 10^{-2}  $ & $(5.2 \pm 0.2) \times 10^{-2}  $ \\
     \bottomrule
  \end{tabular}
\caption{Numerical values of the coupling constants used in the calculations are presented for completeness (see ~\cite{Braun:2000kw,Liu:2008yg,Liu:2009uc,Liu:2010bh} for more details).}
\label{tab:constants}
\end{table}
The masses of the $SU(3)$ partners of $\Omega(2012)$ are obtained in \cite{Polyakov:2018mow} and presented below.
\begin{equation*}
    \begin{aligned}
    m_- =
        \begin{cases}
            1700 \pm 90~\rm{MeV} & \mbox{for~}\Delta ,\\
            1805 \pm 100~\rm{MeV} & \mbox{for~}\Sigma ,\\
            1910 \pm 110~\rm{MeV} & \mbox{for~}\Xi.
        \end{cases}
    \end{aligned}
\end{equation*}
These mass values are used in our numerical analysis. Moreover, for the masses of the ground state baryons, we adapted values from PDG~\cite{PDG:2022pth}.
In addition, the value of the quark condensate is taken as $\qq = -(246_{-19}^{+28}~\rm{MeV})^3$
\cite{Khodjamirian:2011jp}.

The residues of the negative parity $J^P = {3\over2}^-$ baryons are related
with the residues of the radial excitations of the decuplet baryons as follows,
\bea
\label{nolabel}
\lambda_- = \lambda_{rad} \sqrt{m_- - m_+\over m_- + m_+}~.\nnb
\eea 
The residues of radial excitations of the decuplet baryons are calculated in
\cite{Aliev:2016jnp}. Using these results one can easily determine the
residues of the $J^P = {3\over2}^-$ baryons.

The working regions of the Borel mass parameters used in the numerical analysis are
presented in Table \ref{tab2}. Determination of these regions is based on the
criteria that both power corrections and continuum contributions should be
suppressed. The continuum threshold $s_0$ is obtained from the condition
that the mass of the considered states reproduce the experimental values
about 10\% accuracy..


\begin{table}[hbt]
  \renewcommand{\arraystretch}{1.5}
\setlength{\tabcolsep}{6pt}
\begin{adjustbox}{center}
\begin{tabular}{ccccc}
  \toprule
& \multicolumn{2}{c}{Borel mass parameters} & 
  \multicolumn{1}{c}{Continuum threshold }  & 
  \multicolumn{1}{c}{}                          \\
        & $M_1^2~(\rm{GeV}^2)$ &  $M_2^2~(\rm{GeV}^2)$  & $s_0~(\rm{GeV}^2)$    \\
  \midrule
$\Delta \to N \pi$       & $3 \div 4$ & $0.775 \pm 0.025$ & $5.1 \pm 0.1$   \\
$\Sigma \to N K$         & $3 \div 4$ & $0.750 \pm 0.025$ & $5.1 \pm 0.1$   \\
$\Sigma \to \Lambda \pi$ & $3 \div 4$ & $0.750 \pm 0.025$ & $5.1 \pm 0.1$   \\
$\Sigma \to \Sigma \pi$  & $3 \div 4$ & $0.750 \pm 0.025$ & $5.1 \pm 0.1$   \\
$\Xi \to \Lambda K$      & $3 \div 4$ & $0.750 \pm 0.050$ & $6.1 \pm 0.1$   \\
$\Xi \to \Sigma  K$      & $3 \div 4$ & $0.750 \pm 0.050$ & $6.1 \pm 0.1$   \\
$\Xi \to \Xi \pi$        & $3 \div 4$ & $0.750 \pm 0.025$ & $6.1 \pm 0.1$   \\
\bottomrule
\end{tabular}
\end{adjustbox}
  \caption{Working regions of the Borel mass parameters and continuum
threshold $s_0$.}
  \label{tab2}
\end{table}

\begin{figure}[t]
  \centering
  \includegraphics[width=0.75\textwidth]{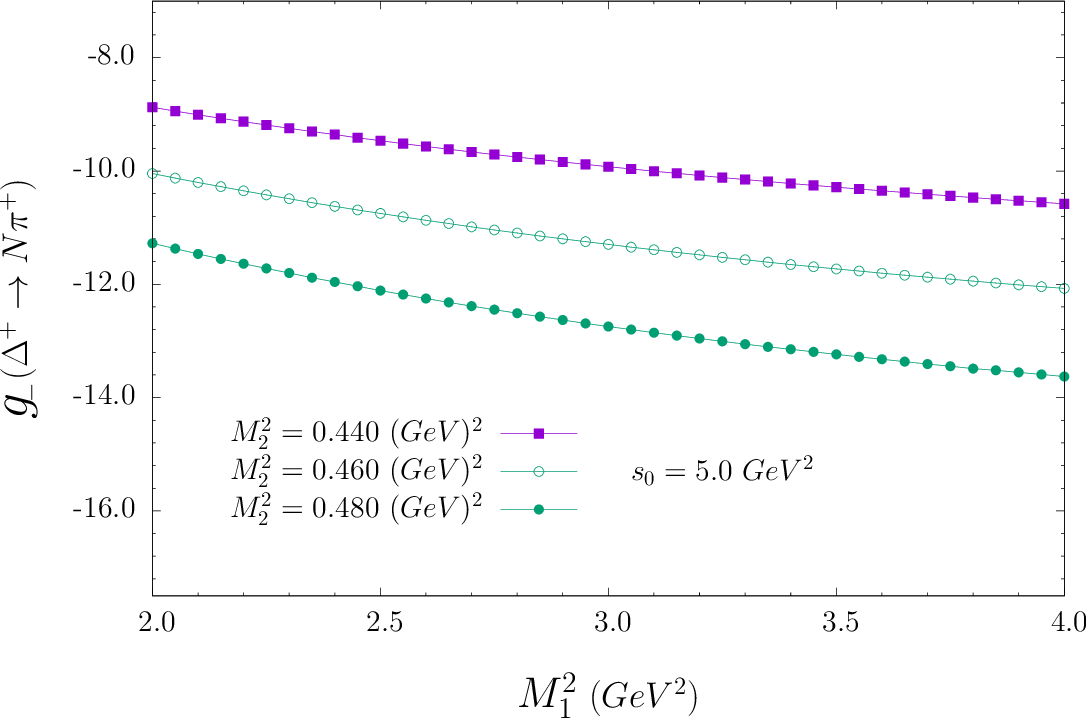}
  \caption{The dependency of the coupling constant of the
$\Delta^+ \to N \pi^+$ transition on the Borel mass 
parameter $M_1^2$, at several fixed values of the Borel parameter $M_2^2$,
and the continuum threshold $s_0=5.0~GeV^2$.}
\label{fig:fig1}
\end{figure}

Having the values of all input parameters at hand, we can proceed to perform
the numerical analysis of the relevant coupling constants. As an example,
in Fig.~\ref{fig:fig1}, we present the dependency of the coupling constant on
$M_2^2$ at the fixed values of the continuum threshold $s_0$ and $M_1^2$
for the $\Delta^+ \to N \pi^+$ transition.
From this figure we observe that there exists good stability of the coupling
constant when $M^2$ varies in its working region (see Table~\ref{tab2}).
The obtained coupling constants are presented in Table \ref{tab3}. The errors in the results for the coupling
constants can be attributed to the uncertainties in the input parameters as well as to the Borel mass parameters $M_1^2$, $M_2^2$, 
and continuum threshold $s_0$.  

After the determination of coupling constants, we can calculate the
corresponding decay channels. Using the matrix elements for the considered $\frac{3}{2}^- \to \frac{1}{2}^+ + \text{pseudoscalar meson}$ transitions, the decay width can be written as,
\bea
\label{eq11}
\Gamma = {g_-^2 \over 24 \pi m_-^2} \Big[(m_- - m_{\cal O})^2 - m_{\cal
P}^2\Big] \vert \vec{p} \vert^3~,
\eea
where
\bea
\label{nolabel}
\vert \vec{p} \vert  = {1\over 2 m_-} \sqrt{m_-^4 + m_{\cal O}^4 + m_{\cal P}^4 -
2 m_-^2 m_{\cal O}^2 - 2 m_-^2 m_{\cal P}^2 -
2 m_{\cal O}^2 m_{\cal P}^2 }~, \nnb
\eea
is the momentum of octet baryon, $m_{\cal O}$ and $m_{\cal P}$ are the mass of the octet baryon and pseudoscalar meson, respectively,
Using the values of the coupling constants obtained within this work, we estimated the decay widths of the relevant transitions that are summarized in Table \ref{tab3}. For comparison we also present the results of the decay widths obtained within frame of the Flavor $SU(3)$ analysis~\cite{Polyakov:2018mow}. We would like to make the following remark at this point. From the expression of the decay width, we see that it is quite sensitive to the mass
splitting among the $SU(3)$ partners of the $\Omega(2012)$ and ground state baryons. Thus, to calculate the coupling constants and decay widths of the transitions under consideration, we used the same masses as in \cite{Polyakov:2018mow}.


\begin{table}[hbt]
  \renewcommand{\arraystretch}{1.5}
\setlength{\tabcolsep}{6pt}
\begin{adjustbox}{center}
\begin{tabular}{cccc}
  \toprule
     Decay channels       & $g_-\,(\rm{GeV}^{-1})$ & $\Gamma\,(\rm{MeV})$ (This work)     & $\Gamma\,(\rm{MeV})$ \cite{Polyakov:2018mow} \\
  \midrule
$\Delta \to N \pi$        & $-11.0 \pm 1.0$   & $48.8   \times (1.0\pm 0.2)$    & 39 - 58 \\
$\Sigma \to N K$          & $-5.6  \pm 0.7$   & $  9.7  \times (1.0\pm 0.3)$    &  7 - 12 \\
$\Sigma \to \Lambda \pi$  & $-7.0  \pm 0.8$   & $ 14.0  \times (1.0\pm 0.3)$    & 11 - 18 \\
$\Sigma \to \Sigma \pi$   & $ 5.5  \pm 0.7$   & $  5.6 \times (1.0 \pm 0.3)$ &  4 -  7 \\
$\Xi \to \Lambda K$       & $-6.9 \pm 1.4$  & $8.0    \times (1.0 \pm 0.3 )$  &  5 - 10 \\
$\Xi \to \Sigma  K$       & $ 6.8  \pm 1.5$  & $4.0    \times (1.0 \pm 0.4 )$  &  2 -  5 \\
$\Xi \to \Xi \pi$         & $ 7.0  \pm 1.1$   & $7.4    \times (1.0\pm 0.3)$    &  5 -  9 \\
\bottomrule
\end{tabular}
\end{adjustbox}
  \caption{Decay widths of the $J^P = {3\over2}^-$ baryons.}
  \label{tab3}
\end{table}

As a final remark, we compare our results with the values obtained within the framework of the flavor $SU(3)$ method~\cite{Polyakov:2018mow}.
In this analysis, the coupling constant for $\Omega \to \Xi K$ is taken as the input parameter, and all the remaining couplings are expressed in terms of this coupling with the help of
$SU(3)$ symmetry relations. Using the experimental value of the decay width $\Omega \to \Xi K$, one can determine the coupling constant of this transition with the help of  Eq. (\ref{eq11}), and hence all the other coupling constants can be determined. 
When we compare our results for the coupling constants and decay widths of the considered decays with those
obtained within flavor $SU(3)$ analysis,  we see they are compatible. Small deviations in the results can be attributed to the $SU(3)$ violation effects.
%
%
\section{Conclusion\label{conclusion}}
In conclusion, we employed the LCSR method to compute the strong coupling constants and decay widths for the $SU(3)$ partners of the $\Omega(2012)$ baryon in $\frac{3}{2}^- \to \frac{1}{2}^+ + \text{pseudoscalar meson}$ transitions. The ``contamination" caused by the $J^P = {3\over2}^+$ baryons are eliminated by considering the linear combinations of the sum rules obtained from different Lorentz structures.
By comparing our decay width results with the findings of \cite{Polyakov:2018mow}, we ascertain the compatibility of our decay width predictions with the outcomes of the flavor $SU(3)$ symmetry analysis. Small discrepancy between the two methods' predictions may be attributed to the $SU(3)$ violation effects. Our results on the  branching ratios can give useful hints about the nature of the $SU(3)$ partners of $\Omega(2012)$ baryon.
%
\section*{Acknowledgments}
We appreciate Y. M. Wang for discussion about numerical analysis. We also thank A. Ozpineci for useful remarks.
\bibliographystyle{utcaps_mod}
\bibliography{all.bib}



\end{document}